\documentclass[twocolappendix]{emulateapj}
\usepackage{graphicx}
\usepackage{amssymb}
\usepackage{amsmath}
\usepackage{natbib}
\newcommand\be{\begin{equation}}
\newcommand\ee{\end{equation}}

\newcommand\ba{\begin{eqnarray}}
\newcommand\ea{\end{eqnarray}}

\begin{document}

\title{On the planetary interpretation of multiple gaps and rings in protoplanetary disks seen by ALMA}

\author{Ryan Miranda\altaffilmark{1,3} and Roman R. Rafikov\altaffilmark{1,2}}

\altaffiltext{1}{Institute for Advanced Study, Einstein Drive, Princeton, NJ 08540}
\altaffiltext{2}{Centre for Mathematical Sciences, Department of Applied Mathematics and Theoretical Physics, University of Cambridge, Wilberforce Road, Cambridge CB3 0WA, UK}
\altaffiltext{3}{miranda@ias.edu}

%%%%%%%%%%%%%%%%%%%%%%%%%%%%%%%%%%%%%%%%%%%%%%%%%%

\begin{abstract}
It has been recently suggested that the multiple concentric rings and gaps discovered by ALMA in many protoplanetary disks may be produced by a single planet, as a result of the complex propagation and dissipation of the multiple spiral density waves it excites in the disk. Numerical efforts to verify this idea have largely utilized the so-called locally isothermal approximation with a prescribed disk temperature profile. However, in protoplanetary disks this approximation does not provide an accurate description of the density wave dynamics on scales of tens of au. Moreover, we show that locally isothermal simulations tend to overestimate the contrast of ring and gap features, as well as misrepresent their positions, when compared to simulations in which the energy equation is evolved explicitly. This outcome is caused by the non-conservation of the angular momentum flux of linear perturbations in locally isothermal disks. We demonstrate this effect using simulations of locally isothermal and adiabatic disks (with essentially identical temperature profiles) and show how the dust distributions, probed by mm wavelength observations, differ between the two cases. Locally isothermal simulations may thus underestimate the masses of planets responsible for the formation of multiple gaps and rings on scales of tens of au observed by ALMA. We suggest that caution should be exercised in using the locally isothermal simulations to explore planet-disk interaction, as well as in other studies of wave-like phenomena in astrophysical disks.
\end{abstract}


%%%%%%%%%%%%%%%%%%%%%%%%%%%%%%%%%%%%%%%%%%%%%%%%%%
%%%%%%%%%%%%%%%%%%%%%%%%%%%%%%%%%%%%%%%%%%%%%%%%%%

\section{Introduction}
\label{sect:intro}

%%%%%%%%%%%%%%%%%%%%%%%%%%%%%%%%%%%%%%%%%%%%%%%%%%

High-resolution observations of protoplanetary disks by ALMA have revealed an exciting richness of ring- and gap-like structures in the spatial distribution of large ($\sim 1$ mm) dust grains on scales of tens of au in a number of systems \citep{ALMA-HL,Andrews-TW,Isella2016,Loomis2017,Andrews-DSHARP}. A remarkable property of these features is their small radial widths --- of order several au ---  clearly distinguishing them from the wide (tens of au) cavities found in transitional disks (e.g., \citealt{Espaillat2014}). 

A number of ideas have been explored for the origin of these features: snowlines --- locations where certain chemical species sublimate \citep{Zhang2015}, zonal flows due to magnetohydrodynamic effects \citep{Flock2015}, and other mechanisms (e.g., \citealt{Takahashi2014}, \citealt{LorenAguilar2015}). All these phenomena are believed to result in small-scale radial variations of gas pressure that lead to dust concentration at the pressure maxima \citep{Whipple1972}, giving rise to the observed axisymmetric gaps/rings. But the most popular (and, probably, the most exciting) explanation for the origin of these features involves planets embedded in disks. 

Since most of the disk features detected by ALMA are narrow, massive (Jupiter-like) planets that carve out wide gaps \citep{LinPap1986,Zhu2011} are unlikely to produce them. In the context of disk-planet interaction, ``massive'' means that the planetary mass $M_{\rm p}$ exceeds the so-called ``thermal mass'' \citep{GR01}
\be
M_\mathrm{th} = \left(\frac{H_{\rm p}}{r}\right)^3 M_* = 1~M_{\rm J} \left(\frac{H/r}{0.1}\right)^3 \frac{M_*}{M_\odot},
\label{eq:Mth}
\ee
at which the perturbation induced by the planetary gravity in the disk is nonlinear from the start; here $r$ is the distance from the central star of mass $M_*$ and $H_{\rm p}$ is the scale height $H = c_{\rm s}/\Omega$ of the disk ($c_{\rm s}$ and $\Omega$ are the sound speed and angular frequency, correspondingly) at the planetary location $r_p$. The multiplicity of narrow gaps/rings may be associated with several lower (sub-thermal) mass planets producing them \citep{Dong2015_gaps,Dipierro2015,Picogna2015,DSHARP_model}, but in some cases this explanation is problematic due to orbital stability arguments \citep{Tamayo2015}. 

At the same time, it is known \citep{R02b} that even a single low-mass planet can carve out multiple gaps. More specifically, nonlinear evolution of the density waves launched by a sub-$M_{\rm th}$ planet converts them into weak shocks relatively close to the planet, a few $H_{\rm p}$ from its orbit \citep{GR01,R02}. Transfer of the wave angular momentum to the disk at these locations carves out two (relatively long-lived) surface density depressions on each side of the planetary orbit \citep{R02b,Duffell,Zhu2013}. The resultant radial pressure perturbations clear two narrow, closely spaced gaps in the dust distribution near the planet (which is located between them), resembling the double gaps seen in sub-mm continuum observations of HL Tau and TW Hya \citep{ALMA-HL,Andrews-DSHARP}. 

While exploring this phenomenon numerically, \citet{DongGaps2017} found that a single low-mass planet can produce not only the two gaps near its orbit but also up to three more narrow gaps in the inner disk closer to the star. \citet{Bae2017} linked the formation of these additional gaps to the nonlinear evolution and shocking of the higher-order spiral arms emerging in the inner disk \citep{Fung2015,BZ18a,BZ18b}. These arms are a generic outcome of linear density wave propagation in disks \citep{Miranda}. 

The idea that a single, relatively low-mass (sub-$M_{\rm th}$) planet can produce a set of narrow gaps/rings over a wide range of distances is, undoubtedly, very interesting. It has been applied by \citet{DSHARP_model}, \citet{Perez2019} and others to explain the multiple narrow features seen in protoplanetary disks by ALMA and to infer the properties of the planets producing them. The goal of our present work is to urge caution regarding the interpretation of observations in terms of characteristics of the putative planets, motivated by the inability of a particular standard tool employed in such studies --- numerical simulations using a locally isothermal equation of state (EoS) --- to properly capture the physics of the planet-disk interaction. 

%%%%%%%%%%%%%%%%%%%%%%%%%%%%%%%%%%%%%%%%%%%%%%%%%%
%%%%%%%%%%%%%%%%%%%%%%%%%%%%%%%%%%%%%%%%%%%%%%%%%%

\section{Statement of the problem}
\label{sect:statement}

%%%%%%%%%%%%%%%%%%%%%%%%%%%%%%%%%%%%%%%%%%%%%%%%%%

Characteristics of multiple gaps/rings produced by the nonlinear evolution of high-order spiral arms excited by a planet \citep{Bae2017} depend primarily on the amount of angular momentum flux (AMF) carried by each high-order density wave \citep{Miranda}. Larger AMF means higher wave amplitude, its faster nonlinear evolution and earlier shocking, shifting the associated axisymmetric feature in the dust distribution closer to the planet \citep{GR01,R02}. The density contrast of the resultant features also scales with the AMF of the waves driving them: transfer of a larger amount of the wave angular momentum to the disk material at the shock causes stronger local perturbation of the gas, and also dust, density. Thus, accurately capturing the AMF behavior of each high-order spiral arm induced by the planet is the key to understanding the properties of the observed gaps/rings, if they are indeed caused by a single planet.

In the absence of dissipation (e.g., at the shock or due to linear damping) the integrated AMF of planet-driven density waves
\be
F_J(r) = r^2 \Sigma(r) \oint u_r(r,\phi) u_\phi(r,\phi) \mathrm{d}\phi
\label{eq:F_J}
\ee
is conserved, i.e. $F_J(r) = $ const ($u_r$, $u_\phi$ are velocity perturbations). At the same time, partitioning of the AMF between the different high-order spiral arms varies with radius \citep{Miranda}.

Given the complexity of the disk-planet interaction \citep{R02,Miranda}, simulations must be used to relate the characteristics of observed axisymmetric features to planetary (mass $M_{\rm p}$ and semi-major axis $r_{\rm p}$) and disk (aspect ratio $H_{\rm p}/r_{\rm p}$) properties  \citep{DongGaps2018}. Because of the numerical costs involved, such simulations usually employ a 2D setup. This is a source of uncertainty, since the planetary torque \citep{DAngelo2010}, as well as wave propagation and dissipation \citep{Lubow1998,Ogilvie1999} may be modified in 3D.

Additionally, and most importantly for our present study, these simulations typically use a {\it locally isothermal EoS} to treat gas thermodynamics \citep{DongGaps2017,Bae2017,DongGaps2018,DSHARP_model,Perez2019,Nazari2019}. This EoS obviates the need to evolve the energy equation and allows a fixed disk temperature profile to be maintained. Its use is often motivated by the expectation of a vertically isothermal structure of externally irradiated protoplanetary disks \citep{Chiang}. However, when the focus is on dynamic features of the flow (such as the density waves), this EoS typically does not provide a good description of thermodynamics of real protoplanetary disks. 

Indeed, one can show that the dynamic response of a gas with adiabatic exponent $\gamma\neq 1$ can be approximated by the isothermal EoS only if the cooling time $t_{\rm c}$ is very short, typically $\Omega t_{\rm c} \ll H/r \sim 0.1$ (Miranda \& Rafikov, in preparation). In protoplanetary disks this regime is realized only at $\gtrsim 80$ au. Thus, the locally isothermal EoS does not accurately represent the physics of disk-planet interaction on scales of several tens of au. 

Moreover, use of this EoS for studying propagation of density waves results in a {\it qualitative bias} stemming from the fact that in locally isothermal disks the AMF of the wave $F_J$ is {\it not conserved}. \citet{Lin2011} and \citet{Lin2015} showed that the AMF of a density wave propagating in locally isothermal disks changes {\it even in the linear regime} due to the torque applied onto the wave by the background shear flow. Instead, in such disks a conserved quantity is\footnote{This result was also stated without a proof in \citet{Lee2016}.} $F_J/c_\mathrm{s}^2$ (Miranda \& Rafikov, in preparation):
\be
\frac{\mathrm{d}}{\mathrm{d}r}\left(\frac{F_J}{c_\mathrm{s}^2}\right) = 0.
\label{eq:cons-iso}
\ee
This reduces to $\mathrm{d}F_J/\mathrm{d}r = 0$ only when $c_{\rm s}$ is constant throughout the disk. 

The difference in the $F_J$ behavior of the density waves between real disks and those modeled using the locally isothermal EoS is very important in light of the aforementioned critical role of the planet-driven wave AMF in determining the characteristics of the gaps/rings observed by ALMA. We now assess the impact of using the locally isothermal EoS for modeling axisymmetric structures in protoplanetary disks.  

%%%%%%%%%%%%%%%%%%%%%%%%%%%%%%%%%%%%%%%%%%%%%%%%%%
%%%%%%%%%%%%%%%%%%%%%%%%%%%%%%%%%%%%%%%%%%%%%%%%%%

\section{Physical and numerical setup}
\label{sect:setup}

%%%%%%%%%%%%%%%%%%%%%%%%%%%%%%%%%%%%%%%%%%%%%%%%%%

%%%%%%%%%%%%%%%%%%%%%%%%%%%%%%%%%%%%%%%%%%%%%%%%%%

\subsection{Basic Disk Model}

%%%%%%%%%%%%%%%%%%%%%%%%%%%%%%%%%%%%%%%%%%%%%%%%%%

We consider the interaction of a planet of mass $M_\mathrm{p}$ on a circular orbit with a radius $r_\mathrm{p}$ and orbital period $t_\mathrm{p} = 2\pi/\Omega_\mathrm{p}$ with a thin disk. The disk initially has a sound speed profile given by 
\be
\label{eq:cs_profile}
c_\mathrm{s}(r) = h_\mathrm{p} r_\mathrm{p}\Omega_\mathrm{p} \left(\frac{r}{r_\mathrm{p}}\right)^{-q/2},
\ee
where $h_\mathrm{p}$ is the disk aspect ratio, $h(r) = H/r = h_\mathrm{p}(r/r_\mathrm{p})^{(1-q)/2}$, evaluated at $r_\mathrm{p}$. We set $h_\mathrm{p} = 0.1$ throughout this letter. We choose $q = 1/2$ (as often assumed for circumstellar disks), but also consider $q = 1$ (corresponding to constant $h(r)$) to assess how the AMF is modified in locally isothermal disks with a more extreme temperature profile.

The initial gas surface density profile is $\Sigma_\mathrm{g}(r) = \Sigma_\mathrm{g,p}\left(r/r_\mathrm{p}\right)^{-1}$ (the value of $\Sigma_\mathrm{g,p} = \Sigma_\mathrm{g}(r_\mathrm{p})$ is arbitrary). This choice does not affect the AMF or the spiral waves in the linear regime. However, the $\Sigma_\mathrm{g}(r)$ profile affects nonlinear dissipation; this dependence will be explored in a future work. 

%%%%%%%%%%%%%%%%%%%%%%%%%%%%%%%%%%%%%%%%%%%%%%%%%%

\subsection{Equation of State}

%%%%%%%%%%%%%%%%%%%%%%%%%%%%%%%%%%%%%%%%%%%%%%%%%%

We consider two different EoS for the gas. The first is the locally isothermal EoS,
\be
P = c_\mathrm{s}^2(r) \Sigma_\mathrm{g},
\ee
where $c_\mathrm{s}(r)$ is a prescribed function of $r$ given by equation~(\ref{eq:cs_profile}). In this case, no energy equation is solved. The second is an ideal EoS,
\be
P = (\gamma - 1) e \Sigma_\mathrm{g},
\ee
where $\gamma$ is the adiabatic index and $e$ is the specific internal energy. In this case, the adiabatic sound speed $c_\mathrm{s}^2 = \gamma(\gamma - 1) e$ is determined by solving the (adiabatic) energy equation, and the $c_s$ profile ~(\ref{eq:cs_profile}) strictly represents only the initial condition for $e$.

To make the most direct comparison possible between the two cases, for the adiabatic EoS we choose $\gamma$ very close to unity, $\gamma = 1.001$. As a result, the disk heats up very slowly and the adiabatic sound speed $(\gamma P/\Sigma_\mathrm{g})^{1/2}$ is nearly identical to the isothermal sound speed $(P/\Sigma_\mathrm{g})^{1/2}$. Therefore, differences between the two cases arise only due to $c_\mathrm{s}$ being either a fixed function of $r$ or a self-consistently evolving variable. Note that the linear response of the disk to the planet is essentially insensitive to the value of $\gamma$ \citep{Miranda}, while the nonlinear wave evolution does depend on $\gamma$ \citep{GR01}. Thus, the use of $\gamma$ very close to unity (and not higher, as would be appropriate for real disks) allows us to focus on the differences in wave propagation arising due to the different AMF behavior between the two chosen EoS.

%%%%%%%%%%%%%%%%%%%%%%%%%%%%%%%%%%%%%%%%%%%%%%%%%%

\subsection{Hydrodynamical Simulations}
\label{sect:hydro}

%%%%%%%%%%%%%%%%%%%%%%%%%%%%%%%%%%%%%%%%%%%%%%%%%%

We perform 2D inviscid hydrodynamical simulations of planet-disk interaction using {\sc fargo3d} \citep{FARGO3d}. We choose a logarithmically-spaced radial grid extending from $r_\mathrm{in} = 0.05 r_\mathrm{p}$ to $r_\mathrm{out} = 5.0 r_\mathrm{p}$, and apply wave damping at $r < 0.06 r_\mathrm{p}$ and $r > 4.5 r_\mathrm{p}$. The planetary potential is softened over a length $0.6 H_\mathrm{p}$. Simulations are performed in pairs, using both of the EoS described previously.

We perform two sets of simulations. In the first set, we choose a high spatial resolution ($N_r \times N_\phi = 3004 \times 4096$, i.e., $65$ cells per $H$ at $r_\mathrm{p}$), and evolve the disk for $\approx 10 t_\mathrm{p}$, sufficient for a quasi-steady perturbation profile to be established across the disk. We use the results of these simulations to characterize the planet-disk interaction through the AMF behavior of planet-induced density waves. We consider planet masses in the range $(10^{-5} - 10^{-3})M_*$ ($\approx 3 M_\oplus - 1 M_\mathrm{J}$ for $M_* = 1 M_\odot$), or $(0.01 - 1) M_\mathrm{th}$ in terms of the thermal mass. 

The second set of simulations uses a lower spatial resolution ($N_r \times N_\phi = 1128 \times 1536$) to allow a much longer evolution timescale, $1000 t_\mathrm{p}$ or more (a few $\times 10^5$ years for $r_\mathrm{p} \approx (30 - 50)$ au). These simulations are used to characterize the long-term evolution of the disk (development of rings/gaps). We choose $M_\mathrm{p} = 0.1, 0.3$ and $1 M_\mathrm{th}$, and consider only a $q = 1/2$ temperature profile.

%%%%%%%%%%%%%%%%%%%%%%%%%%%%%%%%%%%%%%%%%%%%%%%%%%

\subsection{Dust Evolution}
\label{sect:dust}

%%%%%%%%%%%%%%%%%%%%%%%%%%%%%%%%%%%%%%%%%%%%%%%%%%

%%%%%%%%%%%%%%%%%%%%%%%%%
\begin{figure*}
\begin{center}
\includegraphics[width=0.99\textwidth,clip]{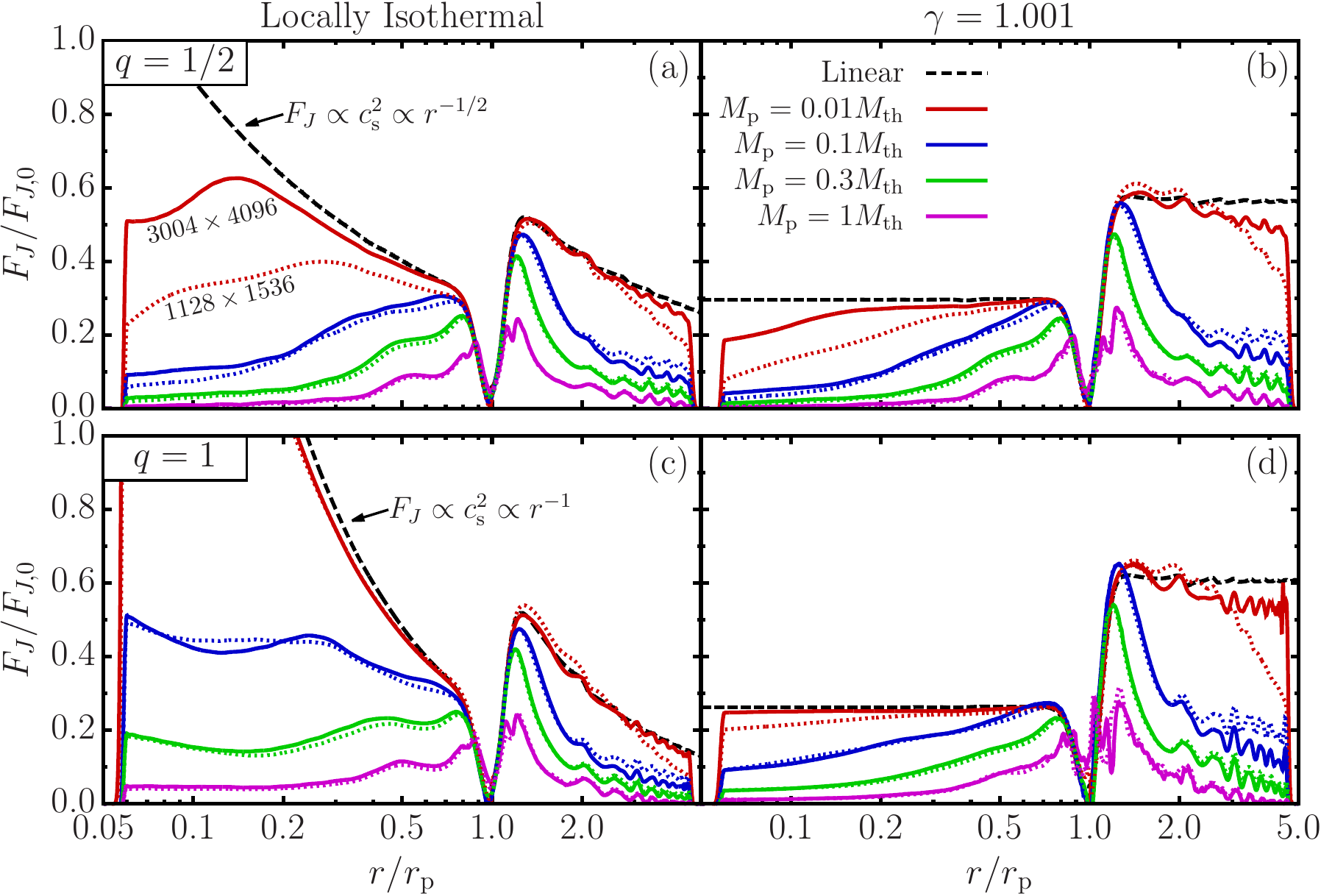}
\caption{Profiles of the planet-induced wave angular momentum flux (AMF) $F_J$ (in terms of the characteristic wave AMF $F_{J,0} = (M_{\rm p}/M_*)^2 h_{\rm p}^{-3}\Sigma_{\rm p} r_{\rm p}^4 \Omega_{\rm p}^2$), at $10 t_\mathrm{p}$ for different planet masses (solid lines), $M_\mathrm{p}$ (expressed in terms of $M_{\rm th}$, see Eq.~(\ref{eq:Mth})), and different temperature profiles (described by the temperature power law index $q$), for locally isothermal (panels (a) and (c)) and adiabatic disks with $\gamma=1.001$ (panels (b) and (d)). Solid lines are the results of high resolution ($N_r \times N_\phi = 1128 \times 1536$) simulations, and dotted lines correspond to the lower resolution ($N_r \times N_\phi = 3004 \times 4096$, as used in our long-term simulations). The black dashed line in each panel is the linear AMF, which, far from the planet, is constant in adiabatic disks but scales as $c_\mathrm{s}^2 \propto r^{-q}$ in locally isothermal disks. Deviations from the linear prediction, more significant for higher $M_{\rm p}$, are caused by nonlinear dissipation.}
\label{fig:amf}
\end{center}
\end{figure*}
%%%%%%%%%%%%%%%%%%%%%%%%%

%%%%%%%%%%%%%%%%%%%%%%%%%
\begin{figure*}
\begin{center}
\includegraphics[width=0.99\textwidth,clip]{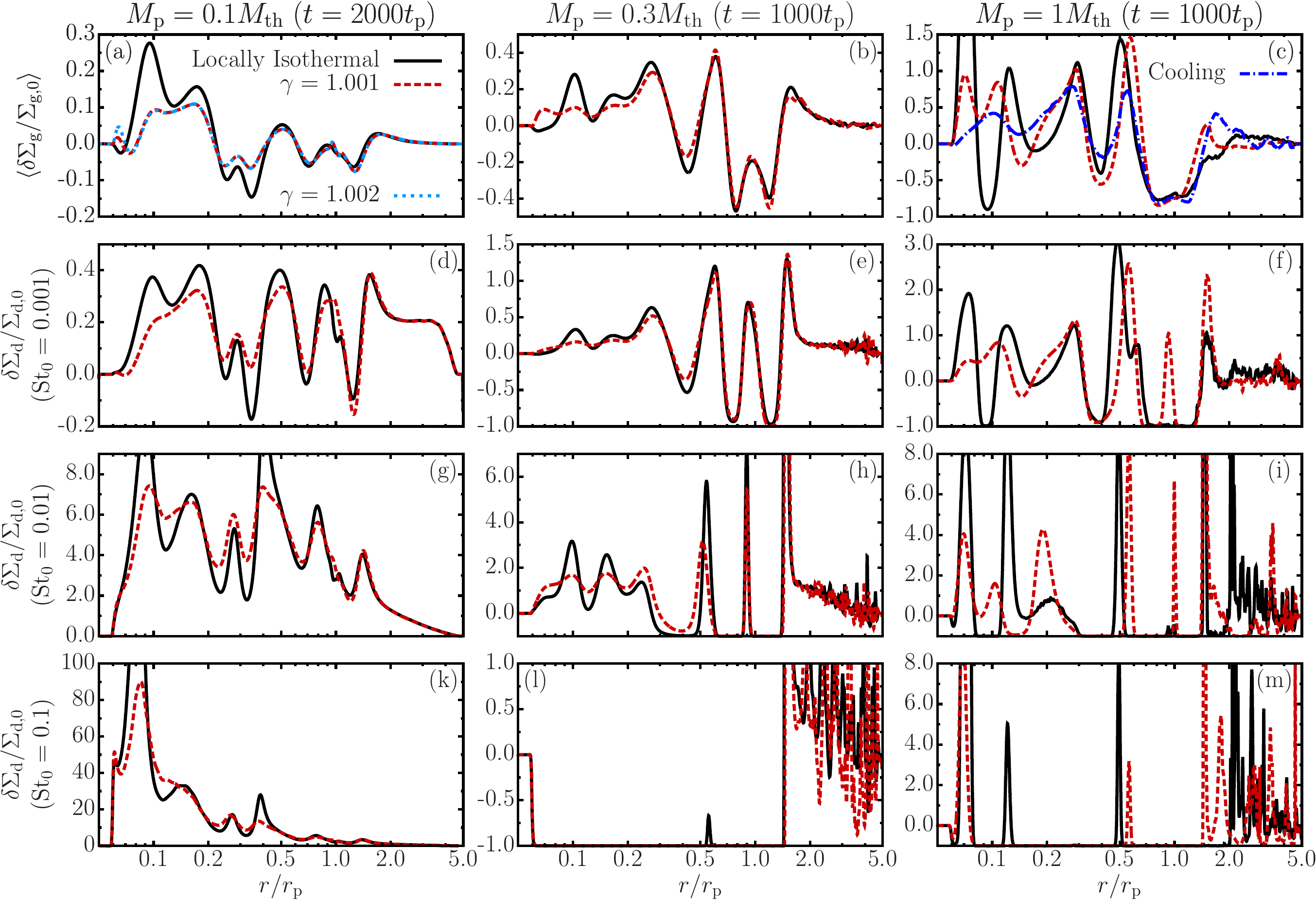}
\caption{Profiles of the fractional perturbations (relative to the initial profiles $\Sigma_{\mathrm{g},0}(r)$ or $\Sigma_{\mathrm{d},0}(r)$) of the azimuthally averaged gas surface density $\langle\Sigma_\mathrm{g}\rangle$ (panels (a)--(c)), and to the surface density of dust with characteristic Stokes number (see eq.~\ref{eq:dustsize}) $\mathrm{St}_0 = 0.001$ ((d)--(f)), $\mathrm{St}_0 = 0.01$ ((g)--(i)) and $\mathrm{St}_0 = 0.1$ ((k)--(m)). Different columns correspond to a different planet masses, $0.1 M_\mathrm{th}$ (left), $0.3 M_\mathrm{th}$ (middle), and $1 M_\mathrm{th}$ (right). The profiles are shown at $1000 t_\mathrm{p}$, except for the case $M_\mathrm{p} = 0.1 M_\mathrm{th}$, which is shown at $2000 t_\mathrm{p}$. Black solid lines show the results for a locally isothermal EoS, which are very different from the red dashed lines --- the results for an adiabatic EoS with $\gamma = 1.001$. In panel (a), the light blue line shows the case with $\gamma = 1.002$ (on top of the red line), and the dark blue line in panel (c) shows the case with $\gamma = 1.001$ and slow cooling. In panels (g)--(m), the vertical scale has been reduced to highlight the differences in the profiles between the different EoS. }
\label{fig:profiles}
\end{center}
\end{figure*}
%%%%%%%%%%%%%%%%%%%%%%%%%

%%%%%%%%%%%%%%%%%%%%%%%%%
\begin{figure*}
\begin{center}
\includegraphics[width=0.80\textwidth,clip]{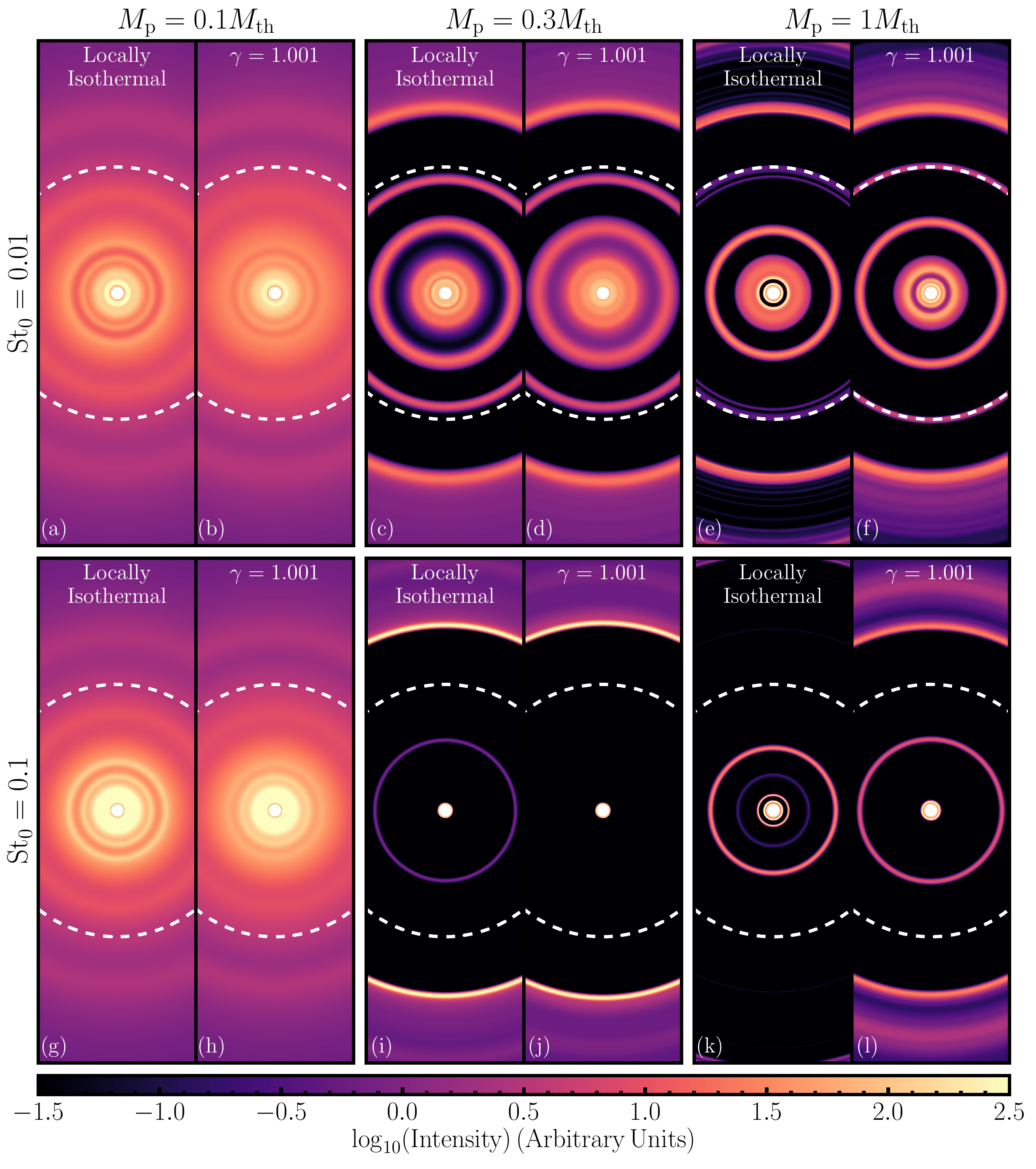}
\caption{Axisymmetric 2D maps of dust continuum emission intensity for different planet masses (columns) and different dust sizes, $\mathrm{St}_0 = 0.01$ (panels (a)--(f)) and $\mathrm{St}_0 = 0.1$ ((g)--(l)). The white dashed circle in each panel indicates the orbit of the planet. Each pair of images shows the emission for the locally isothermal EoS and for the ideal (adiabatic) EoS with $\gamma=1.001$ side-by-side to highlight the differences. The intensity is shown on a logarithmic scale and in arbitrary units.}
\label{fig:intensity}
\end{center}
\end{figure*}
%%%%%%%%%%%%%%%%%%%%%%%%%

We treat dust evolution by post-processing our long-term simulations, using an approximate 1D method. This method neglects the azimuthal structure of the dust, as we are primarily interested in concentric gap/ring structures. It also neglects the dust backreaction, which is equivalent to assuming a low dust-to-gas ratio.

The dust surface density $\Sigma_\mathrm{d}(r)$ obeys the 1D continuity equation,
\be
\label{eq:dusteqn}
\frac{\partial\Sigma_\mathrm{d}}{\partial t} + \frac{1}{r} \frac{\partial}{\partial r}\left(r \Sigma_\mathrm{d} u_{r,\mathrm{d}}\right) = 0.
\ee
Here the radial velocity of the dust is \citep{TL02}
\be
\label{eq:vrd}
u_{r,\mathrm{d}} = \frac{1}{1+\mathrm{St}^2}\left(\overline{u}_{r,\mathrm{g}} + \frac{\mathrm{St}}{\langle\Sigma_\mathrm{g}\rangle\Omega_\mathrm{K}}\frac{\mathrm{d}\langle P\rangle}{\mathrm{d}r}\right),
\ee
where $\langle.\rangle$ indicates the azimuthal average, and $\overline{u}_{r,\mathrm{g}}$is the ``effective'' gas radial velocity which satisfies the 1D continuity equation for $\langle\Sigma_\mathrm{g}\rangle$. In general, $\overline{u}_{r,\mathrm{g}} \neq \langle u_{r,\mathrm{g}}(r,\phi)\rangle$, since $\Sigma_\mathrm{g}(r,\phi)$ satisfies a 2D continuity equation. The Stokes number $\mathrm{St} = \Omega t_\mathrm{s}$, where the stopping time $t_\mathrm{s}$ is the characteristic timescale for aerodynamic drag to change the momentum of a dust particle. In the Epstein drag regime, the Stokes number for a particle with bulk density $\rho_\mathrm{d}$ and size $s_\mathrm{d}$ is
\be
\mathrm{St} = \frac{\pi \rho_\mathrm{d} s_\mathrm{d}}{2\Sigma_\mathrm{g}}.
\ee
Since St varies throughout the disk (due to the variation of $\Sigma_\mathrm{g}$), it is convenient to write
\be
\mathrm{St} = \left(\frac{\Sigma_\mathrm{g}}{\Sigma_\mathrm{g,p}}\right)^{-1} \mathrm{St}_0.
\ee
Therefore, dust dynamics are set by the value of the parameter $\mathrm{St}_0$, which can be related to the particle size and density as
\be
\label{eq:dustsize}
s_\mathrm{d} = 0.64~ \mathrm{mm} ~ \left(\frac{\mathrm{St}_0}{0.01}\right) \left(\frac{\Sigma_\mathrm{g,p}}{10~\mathrm{g~cm}^{-2}}\right) \left(\frac{\rho_\mathrm{d}}{1~\mathrm{g~cm}^{-3}}\right)^{-1} .
\ee

The gas variables $\langle\Sigma_\mathrm{g}\rangle$ and $\langle P \rangle$ are extracted from the hydrodynamical simulations using cubic interpolation (in $r$ and $t$) of snapshots taken every $10 t_\mathrm{p}$. Equation~(\ref{eq:dusteqn}) is solved on a logarithmic grid with $2000$ points, with an adaptive diffusion term added for numerical stability and to prohibit very large contrasts in $\Sigma_\mathrm{d}$. Damping zones relax $\Sigma_\mathrm{d}$ to the initial condition near the grid boundaries. As a result, the total dust mass is not conserved, and typically increases with time due to replenishment near the outer boundary.

%%%%%%%%%%%%%%%%%%%%%%%%%%%%%%%%%%%%%%%%%%%%%%%%%%

\subsection{Emission Maps}

%%%%%%%%%%%%%%%%%%%%%%%%%%%%%%%%%%%%%%%%%%%%%%%%%%

We produce simplified dust continuum emission maps using the computed dust profiles after $(1000 - 2000) t_\mathrm{p}$. The emission is assumed to be optically thin, with intensity $I_\nu(r) = B_\nu[T(r)] \kappa_\nu \Sigma_\mathrm{d}(r)$ at the frequency $\nu$ of ALMA observations, where $B_\nu(T)$ is the Planck function and $\kappa_\nu$ is the dust opacity. In the Rayleigh-Jeans limit, appropriate for mm emission in the outer parts of protoplanetary disks, $B_\nu(T) \propto T \propto r^{-q}$, and so\footnote{For adiabatic disks with $\gamma=1.001$, $T$ deviates negligibly from the initial profile (\ref{eq:cs_profile}).}
\be
I_\nu(r) \propto r^{-q}\Sigma_\mathrm{d}(r).
\ee
Using these assumptions we create pseudo-2D intensity maps using the 1D intensity profile $+$ azimuthal symmetry. These maps assume that all of the emission comes from dust with a single size. These synthetic images, based on an approximate treatment of dust dynamics, serve only to highlight the differences in dust morphology arising in disks evolved with different EoS.

%%%%%%%%%%%%%%%%%%%%%%%%%%%%%%%%%%%%%%%%%%%%%%%%%%
%%%%%%%%%%%%%%%%%%%%%%%%%%%%%%%%%%%%%%%%%%%%%%%%%%

\section{Results}
\label{sect:res}

%%%%%%%%%%%%%%%%%%%%%%%%%%%%%%%%%%%%%%%%%%%%%%%%%%

Radial profiles of the AMF $F_J$ are shown in Fig.~\ref{fig:amf} for several planet masses, temperature profiles, and for the two different EoS, locally isothermal and adiabatic (with $\gamma = 1.001$). The numerical results are shown at $10 t_\mathrm{p}$, for both the low and high resolution simulations described in Section~\ref{sect:hydro}. The AMF profile does not vary much with time as long as the $\Sigma_\mathrm{g}$ profile has not evolved significantly; at later times it is modified by the surface density variations (gaps/rings) produced by the planet. The results of linear calculations are also shown in Fig.~\ref{fig:amf} (dashed lines in each panel). These are computed using the method described in \citet{Miranda}, although in the locally isothermal case, a different master equation must be solved (Miranda \& Rafikov, in preparation). The linear profiles show that far from the planet (outside the wave excitation region) $F_J \approx$ const for adiabatic disks, while $F_J \propto r^{-q}$ for locally isothermal disks, as expected from equation (\ref{eq:cons-iso}). The linear profile for the locally isothermal case with $q = 1$ (Fig.~\ref{fig:amf}(a)) exceeds the vertical scale shown by a factor of three in the inner disk. The numerical results for the smallest planet mass, $M_\mathrm{p} = 0.01 M_\mathrm{th}$, are largely representative of the linear regime.

For the larger planet masses we consider ($0.1, 0.3$, $1 M_\mathrm{th}$), $F_J$ is systematically smaller than the linear value as a result of nonlinear dissipation after the density wave shocks. However, note that for $q = 1/2$, even the $0.01 M_\mathrm{th}$ case shows deviations\footnote{This is seen for $q = 1/2$ but not $q = 1$ as a result of the steeper radial scaling of the wave amplitude dictated by angular momentum flux conservation (see eq.~(16) of \citealt{Miranda}) in the former case. As a result, nonlinear effects accumulate faster in the inner disk for $q=1/2$.} from the linear profile at small radii ($\lesssim 0.1 r_\mathrm{p}$).  Complications due to nonlinear effects aside, by comparing Fig.~\ref{fig:amf}(a) to \ref{fig:amf}(b) or Fig.~\ref{fig:amf}(c) to \ref{fig:amf}(d), we see that $F_J$ is always larger (smaller) in the inner (outer) disk for the locally isothermal EoS as compared to the $\gamma=1.001$ EoS, confirming the general expectation of linear theory. This is true even for a $1 M_\mathrm{th}$ planet, which excites waves that are nonlinear to begin with. Fig.~\ref{fig:amf} also demonstrates that the trend is more pronounced for steeper $T(r)$ profiles (higher $q$).

The high resolution used in the short duration simulations (solid curves in Fig.~\ref{fig:amf}) was chosen to minimize numerical dissipation, ensuring that the quasi-linear behavior of the AMF is captured for low-mass planets. In particular, Fig. \ref{fig:amf}(a),(c) confirm our theoretical expectations and demonstrates the key effect: non-conservation of AMF in locally isothermal disks. However, this effect is clearly also present at the lower resolution of the long term simulations (see the dotted curves in Fig.~\ref{fig:amf}), and thus influences the disk evolution. The lower resolution of the long-term simulations is justified as long as the runs with different EoS use the same resolution (and have all other conditions as identical as possible), so that the results can be directly compared. Fig.~\ref{fig:amf} suggests that the increased resolution does not bring adiabatic and locally isothermal simulations into agreement; in fact, the opposite is true.

Profiles of the azimuthally-averaged gas perturbation $\delta\Sigma_\mathrm{g}$ are shown in Fig.~\ref{fig:profiles}(a)--(c) at $(1000 - 2000) t_\mathrm{p}$ for the long-term simulations with $q = 1/2$. These profiles are time-dependent: the rings/gaps become more pronounced with time. Without explicit viscosity, their density contrasts grow indefinitely. Much of the difference between the locally isothermal and $\gamma=1.001$ cases for the $0.1$ and $0.3 M_\mathrm{th}$ planets results from the differing {\it rate} at which the disk evolves. In the locally isothermal case, the radially-varying AMF of the density waves is higher at small radii, leading to faster disk evolution as they damp.

In general, several (four to six) rings (local maxima of $\Sigma_\mathrm{g}$) as well as a similar number of gaps (local minima of $\Sigma_\mathrm{g}$) are formed. For a $0.1$ or $0.3 M_\mathrm{th}$ planet (Fig.~\ref{fig:profiles}(a)--(b)), the locations of these features are roughly the same for both the locally isothermal and $\gamma=1.001$ disks. This is because the shocking length $l_\mathrm{sh}$ at which the planet-driven spiral waves develop into shocks and drive gap/ring formation is only weakly dependent on the wave AMF: one can show that $l_\mathrm{sh} \propto F_J^{-1/5}(\gamma+1)^{-2/5}$ in the local approximation \citep{GR01}.

However, the amplitudes of the $\Sigma_\mathrm{g}$ features, i.e., the degree of mass variation in them, is significantly larger (often by a factor of several) in the locally isothermal case ($\gamma=1$) than in the $\gamma=1.001$ case. This is particularly evident in the inner disk at $r \lesssim 0.5 r_\mathrm{p}$ and follows from the fact that the gap amplitude is determined by the amount of angular momentum transferred from the wave to the disk, which is proportional to $F_J$. Therefore, a large difference in the wave AMF (resulting from different AMF conservation properties) can lead to a small shift in the gap location (as the value of $\gamma$ is almost the same in our case), while still producing a large difference in amplitude, as indicated by our results.

For the $1 M_\mathrm{th}$ planet (Fig.~\ref{fig:profiles}(c)), gas profiles for the different EoS differ even more significantly. In this case, not only the amplitudes, but also the {\it locations} of the rings and gaps differ between the two cases. One may wonder whether these differences are caused by the planet-induced temperature perturbations\footnote{Although the choice of $\gamma \approx 1$ ensures that there is minimal heating of the disk by shocks, some variation of the temperature profile still occurs in the adiabatic simulation. As the disk evolves, gas initially near the planet is repelled from its orbit, displacing the cooler gas in the outer disk and the hotter gas in the inner disk. This results in an effective advective heating of the outer disk and cooling of the inner disk.}: at $1000$ orbits, $T$ has decreased by $10-20\%$ in the inner disk, and increased by $5-10\%$ in the outer disk for this $M_\mathrm{p}$. In order to assess the role of these temperature variations on our results, we ran a $\gamma=1.001$ simulation with slow cooling, which relaxes $T$ towards the initial profile on a timescale of $t_\mathrm{c} = 1000 \Omega^{-1}$. This keeps the temperature profile much closer to the one used in the locally isothermal simulation, with variations of a few percent at $1000$ orbits. The resulting gas profile is shown by the dot-dashed curve in Fig.~\ref{fig:profiles}(c). The gaps/rings have approximately the same positions as in the purely adiabatic case (although the innermost ring is absent), but their amplitudes are somewhat reduced. However, the profile still better resembles the adiabatic case than the locally isothermal case. Therefore, variations of the temperature profile are not the main driver of the different disk structures found for adiabatic versus locally isothermal disks.

Also shown in Fig.~\ref{fig:profiles} are the radial profiles of the dust density perturbation for particles with different sizes: $\mathrm{St}_0 = 0.001$ (Fig.~\ref{fig:profiles}(d)--(f)), $\mathrm{St}_0 = 0.01$ (Fig.~\ref{fig:profiles}(g)--(i)), and $\mathrm{St}_0 = 0.1$ (Fig.~\ref{fig:profiles}(k)--(m)). These correspond to dust sizes of $0.064$, $0.64$, and $6.4$ mm for a fiducial gas surface density (see eq.~\ref{eq:dustsize}). The profiles for $\mathrm{St}_0 = 0.001$ qualitatively follow those of the gas, although the ring/gap contrasts are enhanced due to radial drift. As a result, differences between the locally isothermal and adiabatic EoS are enhanced in the dust distribution. This is even more evident in the $\mathrm{St}_0 = 0.01$ dust, which is more susceptible to radial drift.

For the largest dust size, $\mathrm{St}_0 = 0.1$, the different gas EoS yield very different dust distributions, especially for the two largest $M_{\rm p}$ we consider. For a $0.3 M_\mathrm{th}$ planet (Fig.~\ref{fig:profiles}(l)), the distribution is primarily distinguished by a ring at $\approx 1.5 r_\mathrm{p}$, with all or most of the dust cleared out inside of this radius, reminiscent of a transition disk. However, while the $\gamma=1.001$ simulation yields a completely cleared cavity, the locally isothermal simulation features an additional narrow dust ring at $\approx 0.5 r_\mathrm{p}$. For a $1 M_\mathrm{th}$ planet (Fig.~\ref{fig:profiles}(m)), the $\mathrm{St}_0 = 0.1$ dust profile has several sharp rings between $0.1 r_\mathrm{p}$ and $2 r_\mathrm{p}$, but their locations and amplitudes are very different for the different gas EoS.

The emission maps for $\mathrm{St}_0 = 0.01$ and $\mathrm{St}_0 = 0.1$ are shown in Fig.~\ref{fig:intensity}. These reflect the same features seen in Fig.~\ref{fig:profiles}. For small $M_{\rm p}$ or dust sizes (Fig.~\ref{fig:intensity}(a)--(b),(c)--(d),(g)--(h)), the strengths of the gaps and rings at $r\lesssim 0.5 r_\mathrm{p}$ are more pronounced for the locally isothermal case compared to the adiabatic case. For larger $M_{\rm p}$ and particle sizes (Fig.~\ref{fig:intensity}(e)--(f),(i)--(j),(k)--(l)), even the presence or absence of some features can depend on the EoS. For example, for $M_\mathrm{p} = 0.3 M_\mathrm{th}$, a faint ring in the dust with $\mathrm{St}_0 = 0.1$ at $\approx 0.5 r_\mathrm{p}$ present in the locally isothermal case is completely absent in the $\gamma=1.001$ case. These images illustrate the significant impact of the density wave AMF non-conservation in the locally isothermal disks on the observable dust emission.

%%%%%%%%%%%%%%%%%%%%%%%%%%%%%%%%%%%%%%%%%%%%%%%%%%

%%%%%%%%%%%%%%%%%%%%%%%%%%%%%%%%%%%%%%%%%%%%%%%%%%
%%%%%%%%%%%%%%%%%%%%%%%%%%%%%%%%%%%%%%%%%%%%%%%%%%

\section{Discussion}
\label{sect:disc}

%%%%%%%%%%%%%%%%%%%%%%%%%%%%%%%%%%%%%%%%%%%%%%%%%%

Our results clearly illustrate the anomalous behavior of the locally isothermal EoS in numerical applications: a tiny ($\mathcal{O}(10^{-3})$) difference in the value of $\gamma$ between the adiabatic $\gamma=1.001$ case and the locally isothermal disk with $\gamma=1$ leads to disproportionately large ($\mathcal{O}(1)$) differences in the outcomes of simulations. A qualitative difference between locally isothermal and adiabatic simulations with $\gamma \approx 1$ was previously pointed out by \citet{Kley2012}, although they studied torque excitation near the planet, which is not strongly affected by AMF non-conservation. To highlight this phenomenon in yet another way we show in Fig.~\ref{fig:profiles}(a) the $\Sigma_{\rm g}$ perturbation computed for an EoS with $\gamma=1.002$, i.e., different from the $\gamma=1.001$ case by the same degree as the locally isothermal case. It is evident that, as expected, $\Sigma_{\rm g}$ perturbations for $\gamma=1.001$ and $\gamma=1.002$ are essentially indistinguishable\footnote{The slight difference between the $\gamma = 1.001$ and $\gamma = 1.002$ profiles at $r \lesssim 0.07 r_\mathrm{p}$ is a numerical boundary effect. It goes away if the inner boundary is placed at a smaller radius.}. This simple exercise emphasizes, in agreement with \citet{Lin2011} and \citet{Lin2015}, that the naive numerical implementation of the locally isothermal EoS (not involving the energy equation) can lead to {\it qualitatively} different results compared to more comprehensive treatments of disk thermodynamics with almost identical basic assumptions (same $T(r)$ profile and $\gamma$ different at the $10^{-3}$ level). Thus, caution must be exercised when interpreting the results of simulations with the locally isothermal EoS.

We emphasize that the goal of this letter is not to reproduce the detailed physics of planet-disk interaction. This would necessarily require a realistic EoS (i.e., $\gamma=7/5$), consideration of cooling/radiative transfer, and a 3D treatment. We merely wish to highlight the anomalies resulting from a locally isothermal assumption due to its non-conservation of AMF for linear waves. Our use of an EoS with $\gamma = 1.001$ (rather than $7/5$) is motivated strictly by the desire to keep the disk $T(r)$ profile as close as possible to the one used in the locally isothermal simulations (a slow cooling used to enforce this condition more strongly does not affect our main conclusions, see Fig.~\ref{fig:profiles}(c)) and to eliminate the effect of varying $\gamma$ on the non-linear wave evolution. Therefore, the differences in the results of our 2D simulations with different EoS can be traced directly to their different AMF conservation properties. Such effects should also arise in 3D simulations, as a result of assumptions made about the disk thermodynamics.

When density waves are damped close to the planet (e.g., due to high viscosity), the anomalous effects caused by adopting the locally isothermal EoS may be less significant. Problems with this EoS arise mainly when waves travel far from the planet, absorbing a significant amount of AMF from the disk flow (see Fig.~\ref{fig:amf}), before depositing the accumulated angular momentum back into the disk at a different location. The exchange of angular momentum between the wave and the locally isothermal disk in the linear regime would also drive anomalous disk evolution near the planet even prior to wave shocking (cf.~\citealt{GN89}). Since the locally isothermal EoS has been widely used in numerical studies of numerous {\it global} phenomena involving waves in disks (e.g., \citealt{Podlewska}, \citealt{Miranda2018}, etc.), some aspects of these problems may need to be reassessed. 

Several authors have used 2D simulations to study multiple rings and gaps produced by planets in low-viscosity disks, and compare them to observed rings/gaps at varying levels of detail \citep{DongGaps2017,DongGaps2018,Bae2017,DSHARP_model,Perez2019,Nazari2019}. All these studies use a locally isothermal EoS with a $q = 1/2$ temperature profile (although \citealt{DongGaps2018} used $q = 1$). The planet masses adopted in these studies fall broadly into the range $(0.1 - 1) M_\mathrm{th}$ (although more massive planets have also been considered), as in our calculations. As a result of using a locally isothermal EoS, not conserving the density wave AMF, modeling efforts such as these may be prone to {\it overestimating} the degree to which a planet sculpts the disk, particularly at small radii ($r \ll r_{\rm p}$). Therefore, the masses of the putative planets responsible for features observed by ALMA may be {\it underestimated} in these studies, especially if features far from the planet are attributed to its influence (e.g., AS 209 system modeled in \citealt{DSHARP_model}). Moreover, since the formation of rings and gaps is a time-dependent process, such studies may also  {\it underestimate the time} required for a planet of a given mass to produce an observed set of rings and gaps. For more massive planets ($\gtrsim 1 M_\mathrm{th}$), discrepancies due to the locally isothermal approximation may be even more significant.

\citet{DongGaps2017}, \citet{Perez2019} and \citet{Nazari2019} also explored the effect of planet migration on the location of rings and gaps produced in the dust distribution. In this regard, we note that the consideration of additional physics such as migration may be premature at this stage, given that the basic gas dynamics of the problem may not have been properly captured by the locally isothermal EoS. 

Although the use of the locally isothermal approximation in numerical studies of planet-disk interaction is ubiquitious, its impact on the density wave dynamics --- AMF non-conservation --- has not yet been fully appreciated. This is perhaps because many studies do not compute AMF, focusing instead on the behavior of the torque density (e.g., \citealt{Arzamasskiy2018}) and phenomena (e.g., vortices) occurring close to the planet (i.e., within a few $H_{\rm p}$). However, the global behavior of the density wave AMF is an excellent indicator of the nonlinear evolution \citep{Dong2011b} as well as other subtle effects \citep{Dong2011a,RP12}. We encourage its broader use in numerical studies.

In a forthcoming study (Miranda \& Rafikov, in prep.) we explore the sensitivity of our results to various disk parameters --- aspect ratio, temperature and density profiles. We also consider a more general disk thermodynamics with $\gamma$ typical for protoplanetary disks and explicit cooling. This setup captures the wave dynamics in a more self-consistent fashion and is preferable to using the locally isothermal EoS in numerical studies.

\acknowledgements

We thank an anonymous referee for the suggestions that helped to improve the paper, and we thank Wing-Kit Lee for useful comments. Financial support of this work was provided by NASA via grant 15-XRP15-2-0139.

\bibliographystyle{apj}
\bibliography{references}

\end{document}